\documentclass[conference]{IEEEtran}
\IEEEoverridecommandlockouts

\usepackage{cite}
\usepackage{amsmath,amssymb,amsfonts}
\usepackage{graphicx}
\usepackage{textcomp}
\usepackage{xcolor}
\def\BibTeX{{\rm B\kern-.05em{\sc i\kern-.025em b}\kern-.08em
    T\kern-.1667em\lower.7ex\hbox{E}\kern-.125emX}}

\usepackage{algorithm}
\usepackage{multirow}
\usepackage{algpseudocode}
\usepackage[misc,geometry]{ifsym}
\algrenewcommand\algorithmiccomment[1]{\hfill{\(//\) #1}}
\usepackage[T1]{fontenc}
\usepackage{pifont}
\usepackage[normalsize]{caption}

\usepackage{enumerate}
\usepackage[shortlabels]{enumitem}

\newcommand{\blue}[1]{\textcolor{black}{#1}}

\newcommand{\cmark}{\ding{51}}%
\newcommand{\xmark}{\ding{55}}%

\DeclareMathAlphabet\mathbfcal{OMS}{cmsy}{b}{n}
    
\begin{document}

\title{
Goal-Oriented Multi-Agent Reinforcement Learning for Decentralized Agent Teams
}

\author{
\IEEEauthorblockN{
    Hung Du\IEEEauthorrefmark{1},
    Hy Nguyen\IEEEauthorrefmark{1},
    Srikanth Thudumu\IEEEauthorrefmark{2},
    Rajesh Vasa\IEEEauthorrefmark{1},
    Kon Mouzakis\IEEEauthorrefmark{1}}
\IEEEauthorblockA{
    \IEEEauthorrefmark{1}\textit{Applied Artificial Intelligence Initiative (A2I2), Deakin University, Geelong, VIC, Australia}\\
    \texttt{\{hung.du,hy.nguyen,rajesh.vasa,kon.mouzakis\}@deakin.edu.au}\\[0.2ex]   
    \IEEEauthorrefmark{2}\textit{Institute of Applied Artificial Intelligence and Robotics (IAAIR), Germantown, TN, USA}\\
    \texttt{srikanth@iaair.ai}}
}

\maketitle

\begin{abstract}
Connected and autonomous vehicles across land, water, and air must often operate in dynamic, unpredictable environments with limited communication, no centralized control, and partial observability. These real-world constraints pose significant challenges for coordination, particularly when vehicles pursue individual objectives. To address this, we propose a decentralized Multi-Agent Reinforcement Learning (MARL) framework that enables vehicles, acting as agents, to communicate selectively based on local goals and observations. This goal-aware communication strategy allows agents to share only relevant information, enhancing collaboration while respecting visibility limitations. We validate our approach in complex multi-agent navigation tasks featuring obstacles and dynamic agent populations. Results show that our method significantly improves task success rates and reduces time-to-goal compared to non-cooperative baselines. Moreover, task performance remains stable as the number of agents increases, demonstrating scalability. These findings highlight the potential of decentralized, goal-driven MARL to support effective coordination in realistic multi-vehicle systems operating across diverse domains.

\end{abstract}


\section{Introduction}

Recent advances show that sophisticated AI agents can solve complex tasks and achieve human-like performance in certain contexts \cite{du2024survey}. However, single agents face limitations in scalability, adaptability, and reliability. While parallelization can speed up task execution, it does not enable agents to tackle more complex tasks that require specialization \cite{amirkhani2022consensus}. To overcome these limitations, multi-agent system architectures have emerged, where agents communicate and coordinate to handle complex, dynamic environments—often leveraging Multi-Agent Reinforcement Learning (MARL) to manage interaction dynamics.

In MARL, an agent communicates and interacts with other agents within the same environment. This supports the agent in making decisions based both on its own understanding of the world and on its observations of the actions taken by the other agents. Often, a naive design tactic is applied allowing open communication between all agents which generates a large amount of information within the environment. This forces us to provision an environment with sufficient bandwidth, low latency, and high compute. The core challenge, however, is the requirement for the agent to have a smart filter that can assess the value of information against the goal and that which assists with coordination. Addressing this challenge requires agents to adopt a communication strategy \blue{and coordination} that contextually determines situations. 

\blue{Communication is the process of creating a medium for agents to exchange information, whereas coordination focuses on retrieving, sharing, and combining that information to accomplish specific tasks. Existing strategies can be classified into three categories: Centralized Training and Centralized Execution (CTCE), Centralized Training and Decentralized Execution (CTDE), and Decentralized Training and Decentralized Execution (DTDE). CTCE strategies train all agents using a shared, centralized critic with access to global information, aiming to optimize coordination. During execution, a centralized policy with global observations directly controls all agents. However, in practical scenarios, agents often need to act independently based on local observations. CTDE strategies \cite{gupta2017cooperative,rashid2020monotonic,xu2021learning,ruan2022gcs,pesce2023learning,nayak2023scalable} address this by developing decentralized policies for execution while leveraging a centralized critic during training. These strategies assume (i) agents share a common goal, enabling the use of a centralized critic to evaluate decentralized policies, and (ii) agents can communicate and coordinate directly at every time step. Despite their advantages, CTDE strategies yield suboptimal policies in many real-world scenarios where agents have individual goals and limited observability of others' behavior. DTDE strategies \cite{tan1993multi,de2020independent,jin2022v,daskalakis2023complexity,skrynnik2024learn} tackle these limitations by enabling agents to operate in fully decentralized settings where local observations and knowledge are utilized to optimize their objectives. While DTDE agents can be more robust and adaptable to uncertainties, they face two significant challenges: (i) exhaustive exploration, and (ii) inefficient sharing of experience and knowledge. This can be attributed to the absence of central coordination, restricted observability among agents, and increasing number of agents entering the environment.}

\begin{figure*}[!ht]
    \centering
    \includegraphics[width=1\linewidth]{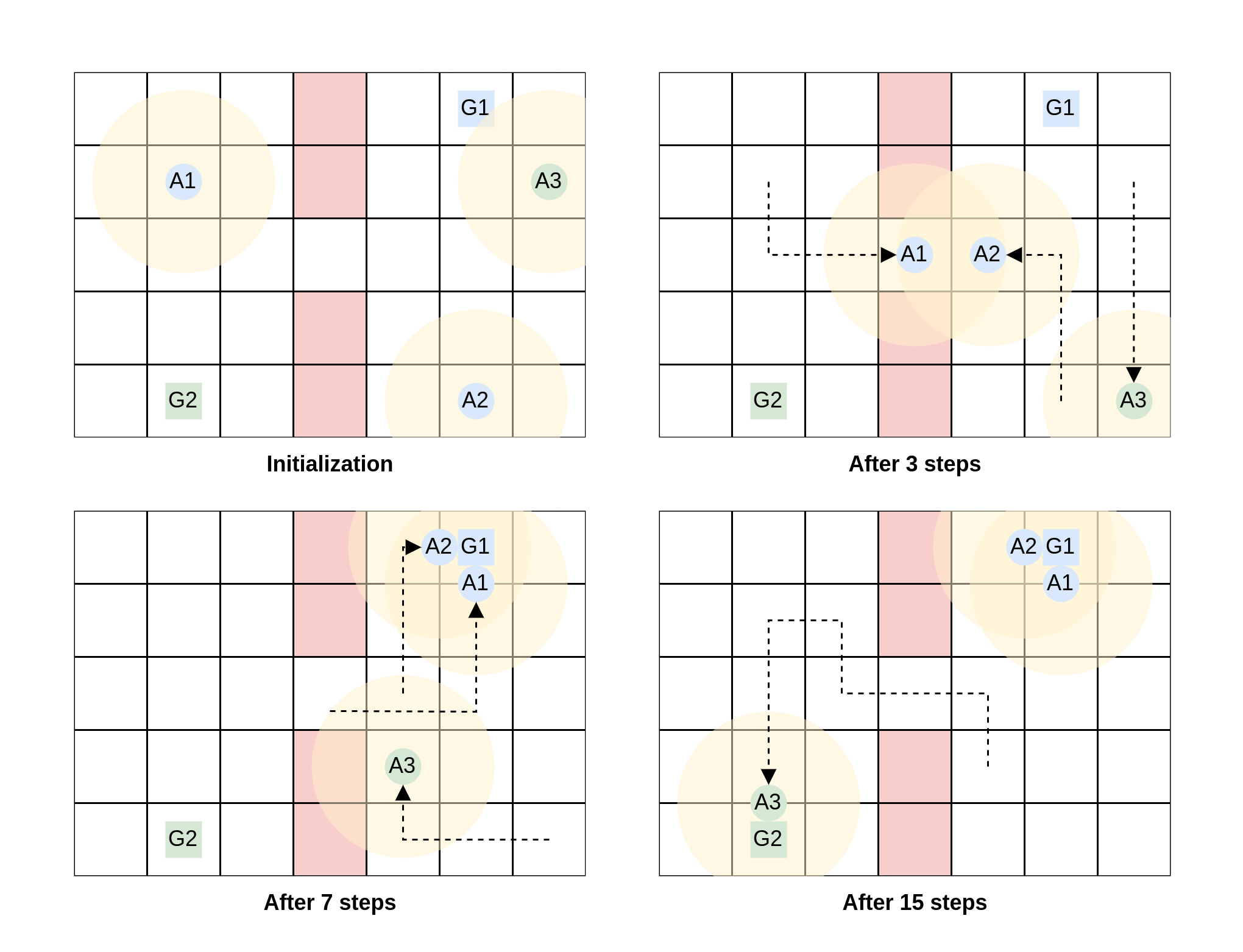}
    \caption{The illustration of our coordination strategy. 
    Agents begin at fixed positions. Agents 2 and 3 do not coordinate upon encountering each other due to differing goals. At step three, Agents 1 and 2 meet and coordinate, reaching their goals with four extra steps, while Agent 3, acting independently, takes 12 additional steps.}
    \label{fig:overview}
\end{figure*}

\blue{To overcome the challenges in DTDE strategies, a naive approach is to enable agents to share their local observations, which can be used to optimize their policies toward individual goals \cite{zhang2019efficient,jiang2022i2q}.
However, this approach often introduces a substantial amount of irrelevant information relative to an agent's goal. This can increase learning complexity and degrade performance. While several approaches have been proposed to address these issues \cite{yuan2022multi,ba2024cautiously}, these approaches often focus on optimizing agents toward a shared system goal. This leads to ineffective communication and coordination when agents pursue individual objectives. To resolve this, it is crucial to incorporate agents' awareness of individual goals into their communication and coordination processes.} In this paper, we propose an \blue{MARL} approach \blue{in fully decentralized settings} where: (1) each agent has its own goal and limited observability of other agents' behavior; \blue{and} (2) an agent communicates and coordinates with other agents \blue{if they share the same goal (see also Figure \ref{fig:overview})}. For our experiments, we focus on multi-agent navigation where agents cooperate to navigate towards their respective goals in \blue{complex} grid environments \blue{with obstacles}. \blue{Our evaluation demonstrates that goal-aware communication and coordination under restrictive conditions enhance overall performance and success rates, outperforming both non-collaborative agents and those employing unrestricted communication and coordination strategies.} The remainder of the paper is organized as follows: Section \ref{sec:rw} reviews related work; Sections \ref{sec:problem} and \ref{sec:approach} present the problem formulation and our method; Section \ref{sec:exp} reports experimental results; and Section \ref{sec:concl} concludes.

\section{Related Work} \label{sec:rw}

In Multi-Agent Reinforcement Learning (MARL), early work by \cite{tan1993multi} showed that agents using Independent Q-Learning (IQL) in cooperative settings can outperform fully independent agents. Given that agents operate based on local observations and make decisions independently, MARL problems are often modeled as Decentralized Partially Observable Markov Decision Processes (Dec-POMDPs) \cite{oliehoek2016concise} (see Section \ref{sec:problem}). To stabilize training, the Centralized Training with Decentralized Execution (CTDE) paradigm \cite{gupta2017cooperative} has been widely adopted, where agents use shared global information and a centralized critic during training, but execute policies independently. Recent CTDE-based approaches \cite{rashid2020monotonic,xu2021learning,nayak2023scalable,pesce2023learning} have demonstrated strong coordination but assume agents share a common goal and can share experiences—assumptions that often break down in real-world settings with heterogeneous objectives and limited communication. Fully decentralized methods 
\cite{tan1993multi,jin2022v,yuan2022multi,daskalakis2023complexity,skrynnik2024learn,ba2024cautiously}
remove these assumptions by training agents with local critics, but face two major challenges: inefficient exploration and limited knowledge sharing. To overcome these, we propose a fully decentralized algorithm that integrates agents’ individual goal awareness into their communication and coordination strategies.

\blue{Facilitating communication among agents is crucial for addressing the challenge of exhaustive exploration in fully decentralized MARL algorithms.} Attentional Communication Model (ATOC) \cite{jiang2018learning} was proposed to encode each agent's local observations, aggregates such representations and utilizes the aggregated information to instruct the agent when to communicate. Extending this, \cite{liu2020who2com} developed a framework that combines agents’ observations to facilitate the selection of agents during communication. In addition, \cite{liu2020when2com} applied graph-based estimation to enable agents to form communication groups and determine the events of communication. \blue{Furthermore, considerable research efforts \cite{bohmer2020deep,pesce2023learning,ba2024cautiously}
have focused on developing filtering mechanisms to optimize communication among agents. However, these mechanisms are designed within the CTDE strategy, making them incompatible with fully decentralized MARL algorithms. Our approach differs from these approaches in two key aspects: (i) agents operate in fully decentralized settings, and (ii) agents engage in restrictive communication that incorporates goal awareness.}

Effective coordination is critical in fully decentralized MARL, where knowledge sharing is inherently limited. Value-Decomposition Networks (VDN) \cite{sunehag2017value} enable agents to learn from joint actions but struggle to select optimal strategies in decentralized settings. QMIX \cite{rashid2020monotonic} addressed this by using a monotonic value function to align local and global value functions. To capture inter-agent relationships, MGAN \cite{xu2021learning} employed Graph Convolutional Networks (GCNs).
Graph-based Coordination Strategy (GCS) \cite{ruan2022gcs} further modeled team policies via graph representations, and \cite{pesce2023learning} proposed a state-dependent communication graph to regulate information flow. While these methods show strong coordination, they are all designed within the Centralized Training with Decentralized Execution (CTDE) framework, making integration into fully decentralized settings challenging. 
Building on the concepts of transfer learning \cite{zhuang2020comprehensive} and federated learning \cite{zhang2021survey}, our method employs weight merging to consolidate knowledge among agents with aligned goals, enabling efficient coordination without violating decentralization constraints.

\section{Problem Preliminary} \label{sec:problem}
In our approach, multiple agents make decisions independently, each with different observations. This approach is therefore modeled as a decentralized partially observable Markov decision process (Dec-POMDP) \cite{oliehoek2016concise} defined by the following tuple: $(n, \mathcal{S}, \{\mathcal{A}_i\}_{i=1}^{n}, T, \{\mathcal{R}_i\}_{i=1}^{n}, \{\mathcal{O}_i\}_{i=1}^{n}, P, \gamma)$. Here, $n$ represents the number of agents, $\mathcal{S}$ is the set of states, $\{\mathcal{A}_i\}_{i=1}^{n}$ denotes the set of action sets for each agent, $T: \mathcal{S} \times \mathcal{A}^n \rightarrow \mathcal{S}'$ is the state transition probability function following the joint actions $\mathcal{A}^n = (a_{1}, a_{2}, \ldots, a_{n})$, $\{\mathcal{R}_i\}_{i=1}^{n}$ is the set of rewards for each agent, $\{\mathcal{O}_i\}_{i=1}^{n}$ represents the set of observations for each agent, $P: \mathcal{S} \times \mathcal{A}^n \rightarrow \mathcal{O}'$ is the observation probability function, and $\gamma \in [0, 1]$ is the discount factor.
\blue{We define a system goal consists of $m$ individual goals, denoted as $\mathcal{G} = \{g_1, \ldots, g_m\}$.}
An agent is initialized with \blue{an individual} goal. \blue{If the agent accomplishes its goal, it will stay in the same position.} In addition, the agent has an observation range and only communicate with other agents that are within the range. \blue{Our approach allows an agent to share its learning weights and obtain others' learning weights if they have the same goal.}
In addition, each independent agent utilizes an actor-critic framework \cite{peters2008natural} to select the optimal action at each time step. The loss functions for the actor and critic are estimated separately as follows:
\blue{
\begin{equation} \label{eq:actor_loss}
    \mathcal{L}_{\text{actor}}(\theta^{\mu}) = -\mathbb{E}[\log{\pi_{\theta}(a|s)A^{\pi}(s, a)}]
\end{equation}
\begin{equation} \label{eq:critic_loss}
\mathcal{L}_{\text{critic}}(\theta^{w}) = \mathbb{E}\left[\left(R(s, a, s') + \gamma V^{\pi}(s'; \theta^{w}) - V^{\pi}(s; \theta^{w})\right)^{2}\right]
\end{equation}
}

\section{Our Approach} \label{sec:approach}
Our approach aims to enhance the learning and exploration processes \blue{agents in the fully decentralized settings. To achieve this, we equip agents with goal-aware capabilities for communication and coordination. For experimental purposes, we design our approach within complex grid environments containing obstacles. 
}

\subsection{Environment and Rewards}
We construct a grid environment denoted by $\text{M}^{\text{w} \times \text{h}}$ (see also Figure \ref{fig:overview}). This environment contains \blue{the following entities: $n$ agents, $m$ objects and $k$ obstacles. The position of an entity is represented by $(x, y)$.} An agent's goal, denoted by $g = (x, y)$, is the position of an object that the agent aims to move towards. At each time step, the current state of an agent is represented by the agent's current position: $s^{t}_{i} = (x^{t}_{i}, y^{t}_{i})$. An agent can choose from five possible actions: staying, moving up, moving down, moving left, or moving right within the boundaries of the environment. \blue{In addition, an agent cannot move to cells occupied by obstacles.} Multiple agents can occupy the same cell. An agent's task is considered complete if it reaches its goal and remains in that position. The sparse reward function of an agent is defined as:
\begin{equation} \label{eq:reward}
    R(s^{t}_{i}) = \begin{cases}
        1 & \text{if } s^{t}_{i} = g_{i} \\
        -\lambda_{\text{stay}} & \text{if } \left( s^{t - 1}_{i} = s^{t}_{i}  \right) \land \left( s^{t}_{i} \neq g_{i} \right) \\
        \blue{\frac{1}{\Delta\left(s^{t}_{i}, g_{i}\right)}} & \text{if } \left(s^{t - 1}_{i} \neq s^{t}_{i} \right) \\
        -1 & \text{otherwise} \\
    \end{cases}
\end{equation}
The reward value ranges between -1 and 1. An agent receives a reward of 1 if its position matches its goal. If the agent remains in a cell that is not its goal, it is penalized by $\lambda_{\text{stay}} \in (0, 1)$. To incentivize movement towards the goal, an agent receives a reward of \blue{$\frac{1}{\Delta\left(s^{t}_{i}, g_{i}\right)}$ where $\Delta > 0$} is the geometric distance between $s^{t}_{i}$ and $g_{i}$. \blue{This indicates} that the closer the agent is to the goal, the higher the reward it receives.

\begin{table}[!t]
    \centering
    \begingroup
    \scriptsize
    \renewcommand{\arraystretch}{1.5}
    \begin{tabular}{|c|c|c|c|c|}
        \hline
        \textbf{Type of Agent} & \multicolumn{2}{|c|}{\textbf{Collaboration}} & \multicolumn{2}{|c|}{\textbf{Observation Range}} \\
        \hline
        & Unrestricted & Goal-aware & Unrestricted & Limited \\
        \hline
        A1 & N/A & N/A & N/A & N/A \\
        A2 & \cmark & \xmark & \cmark & \xmark \\
        A3 & \cmark & \xmark & \xmark & \cmark \\
        A4 & \xmark & \cmark & \cmark & \xmark \\
        A5 & \xmark & \cmark & \xmark & \cmark \\
        \hline
    \end{tabular}
    \endgroup
    \caption{Agent types characterized by collaboration and observation range.}
    \label{tab:agent_feat}
    \vspace{-16pt}
\end{table}

\subsection{State, Action and Relay Buffer}

Each agent possesses its own actor-critic \blue{framework}. The actor's goal is to choose the optimal action based on the agent's current state, while the critic's role is to evaluate the state-action pair. Similar to the Deep Deterministic Policy Gradient (DDPG) algorithm \cite{lillicrap2015continuous}, we utilize deep neural networks in both the actor and the critic to model the state and action. Additionally, each agent has its own memory, known as the relay buffer $\mathcal{B}$, which stores up to $H$ experiences of the agent. An experience consists of the tuple $\left( s^{h}_{i}, a^{h}_{i}, r^{h}_{i}, s^{h+1}_{i} \right)$ from past interactions.

The actor network of the $i^{\text{th}}$ agent, denoted by \blue{$\mu_{i}(s|\theta_{i}^\mu)$}, is initialized with random weights and parameterized by $\theta_{i}^\mu$. Given the current state of the agent \blue{$s$}, the network aims to generate the weight distribution for five actions, \blue{denoted as} $z$. \blue{Note that $s$ consists of the agent position and the index of its individual goal $\mathcal{I}_g$, making $s = (x, y, \mathcal{I}_g)$. We use $\mathcal{I}_g$ to incorporate goal semantics into the agent's action selection process. For simplicity, we adopt the concatenated architecture outlined in \cite{schaul2015universal}. The distribution $z$} is then converted into the probability distribution as follows:
\begin{equation}
    \sigma(z_{l}) = \frac{e^{z_{l}}}{\sum_{k=1}^{K}{e^{z_{k}}}}
\end{equation}
where $l, k \in K$ are indices of actions. The exploration-exploitation dilemma is commonly controlled by the use of $\epsilon$ with a specific threshold depending on the task setting.
However, the choice of $\epsilon$ is not robust because it varies across scenarios. To address this challenge, we apply the multinomial sampling on the probability distribution of actions. This aims to ensure two facets: (1) all actions have a chance to be selected; and (2) an action with the high probability will be more likely to be selected. \blue{To enhance exploration and optimize action selection, an entropy regularization term is incorporated into the actor network parameters \cite{mnih2016asynchronous}.
Equation \ref{eq:actor_loss} is then modified as:
\begin{equation} \label{eq:actor_loss_2}
    \mathcal{L}_{\text{actor}}(\theta^{\mu}) = -\mathbb{E}[\log{\pi_{\theta}(a|s)A^{\pi}(s, a)} + \beta H(\pi_{\theta}(\cdot|s))]
\end{equation}
where $H$ is the entropy, $\beta \in [0, 1]$ is the entropy coefficient that controls how much to prioritize exploration. While the high value of $\beta$ favors exploration, the low value of $\beta$ favors exploitation.}

The critic network of the $i^{\text{th}}$ agent, denoted by \blue{$Q_{i}(s, a|\theta_{i}^Q)$}, is also initialized with random weights and parameterized by $\theta_{i}^Q$. Given the state \blue{with the goal semantics} and the corresponding selected action, the network aims to generate a value that can be utilized to evaluate \blue{the quality of that action}.

\subsection{Coordination Strategy}

\blue{An agent communicates and coordinates with others within its observation range}, illustrated in Figure \ref{fig:overview}. This range consists of $C$ cells surrounding the agent's current position and within the environment boundaries. The range is denoted by $c \in \mathbb{Z}^{+}$. During \blue{the} communication phase, the agent shares its goal and identifies other agents with the same objective \blue{(i.e., peers)}. Instead of exchanging entire historical experiences, which can be costly, agents with the same goal share their knowledge through \blue{the} weight sharing \blue{mechanism} as follows:
\blue{
\begin{equation}
    \theta^{Q}_{i} = (1 - \alpha) \theta^{Q}_{i} + \alpha \frac{1}{K}\sum_{j = 0}^{K}{\theta^{Q}_{j}}
\end{equation}
\begin{equation}
    \theta^{\mu}_{i} = (1 - \alpha) \theta^{\mu}_{i} + \alpha \frac{1}{K}\sum_{j = 0}^{K}{\theta^{\mu}_{j}}   
\end{equation}
}
where $K \leq N$ is the number of peers within the \blue{observation} range, \blue{and $\alpha \in [0, 1]$ is the dampening factor that balance the agent's parameters with those aggregated from its peers. To minimize the substantial influence of an agent's peers on its learning weights, we suggest keeping $\alpha$ as small as possible.}

\section{Experiments} \label{sec:exp}

\blue{In this study, we propose a novel communication and coordination strategy to improve the task performance of decentralized agents. Since our approach operates in fully decentralized settings, comparisons with existing CTDE approaches fall outside the scope of this work. Instead, we conducted ablation studies to examine the performance improvements of agents trained using our method. Details of our experiments are provided below.}

\subsection{Experiment Details} \label{sec:exp_d}

\begin{table}[!t]
    \centering
    \begingroup
    \scriptsize
    \renewcommand{\arraystretch}{1.5}
    \begin{tabular}{|c|c|c|c|c|c|c|}
        \hline
        \textbf{Environment} & \textbf{No.} & \textbf{Types of Agent} & $\boldsymbol{N}$ & $\boldsymbol{G}$ & $\mathbfcal{E}$ & $\mathbfcal{T}$ \\
        \hline
        \multirow{2}{*}{small} & 1 & A1, A2, A3, A4, A5 & 3 & 2 & 2500 & 400 \\
        & 2 & A1, A5 & 4 & 2 & 2500 & 400 \\
        \hline
        large & 3 & A1, A5 & 10 & 2 & 400 & 2500 \\
        \hline
    \end{tabular}
    \endgroup
    \caption{Summary of scenarios conducted to evaluate our approach. Here, $N$ represents the number of agents in the environment, $G$ denotes the number of goals, and $\mathcal{E}$ indicates the number of episodes. Furthermore, we set values for $\mathcal{E}$ and $\mathcal{T}$ such that $\mathcal{E} \times \mathcal{T} = 10^6$.}
    \label{tab:exp_track}
    \vspace{-16pt}
\end{table}

\begin{figure}[!t]
    \centering
    \includegraphics[width=0.85\linewidth]{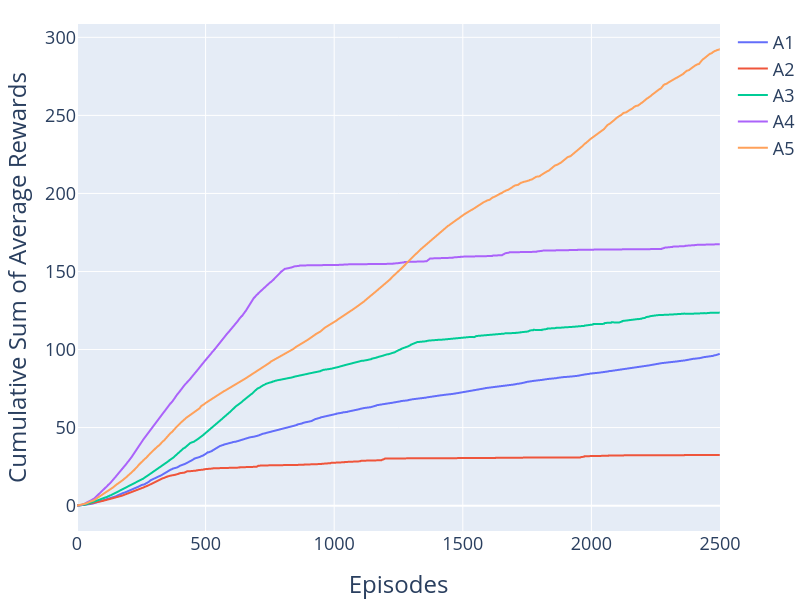}
    \caption{Comparison between agent types in Scenario 1.}
    \label{fig:s1}
    \vspace{-6pt}
\end{figure}

\begin{table}[!ht]
    \centering
    \begingroup
    \scriptsize
    \renewcommand{\arraystretch}{1.5}
    \begin{tabular}{|c|c|c|c|}
        \hline
        \textbf{Types of Agent} & \textbf{Agent 1} & \textbf{Agent 2} & \textbf{Agent 3} \\
        \hline
        A1 & 243 $\pm$ 93 & 178 $\pm$ 103 & 166 $\pm$ 92 \\
        A2 & 214 $\pm$ 101 & 223 $\pm$ 114 & 179 $\pm$ 104 \\
        A3 & 219 $\pm$ 94 & 174 $\pm$ 101 & 171 $\pm$ 96 \\
        A4 & 126 $\pm$ 83 & 101 $\pm$ 81 & 166 $\pm$ 92 \\
        A5 & 171 $\pm$ 93 & 136 $\pm$ 93 & 166 $\pm$ 92 \\
        \hline
    \end{tabular}
    \endgroup
    \caption{The average number of steps taken by each agent during successful episodes in Scenario 1.}
    \label{tab:s1}
\end{table}

\blue{We evaluate our approach in complex grid environments of sizes $\text{M}^{10\times10}$ (small) and $\text{M}^{20\times20}$ (large), which contain obstacles.}
\blue{In our experiments, we designed five types of agents based on their collaboration and observation ranges (see also Table \ref{tab:agent_feat}). These two features are created for collaborative agents and not applicable to non-collaborative agents (A1), which perform tasks independently. Collaboration is categorized into two types: unrestricted collaboration and goal-aware collaboration. In unrestricted collaboration, an agent can communicate and coordinate with any agent in the environment, while in goal-aware collaboration, interaction is limited to agents sharing the same goal. Furthermore, agents may have either an unrestricted or limited observation range. The unrestricted range enables an agent to collaborate with all agents in the environment, regardless of their positions. Meanwhile, the limited range restricts collaboration to agents within the agent's observation range.}
\blue{We designed three scenarios, as outlined in Table \ref{tab:exp_track}. Each type of agent was evaluated independently. In addition, one of our objectives is to determine the best-performing agent type for each scenario. Detailed descriptions of each scenario are provided below:}
\begin{enumerate}
    \item This scenario involves three agents: two agents pursuing the same goal (e.g., $g_1$) and one agent pursuing a different goal (e.g., $g_2$). The two main objectives are: (i) validating whether collaborative agents (A2 through A5) achieve better performance than independent agents (A1), and (ii) identifying the best-performing collaborative agent type.
    \item This scenario consists of two teams of agents, each involving two agents pursuing the same goal. Our experiment showed that A5 outperforms the other agent types (see Section \ref{sec:r_n_d}), and hence, we focus on A5 in this scenario. The goal of this scenario is to evaluate whether collaborative teams can reduce the number of steps each agent takes to complete the task and enhance overall system performance.
    \item This scenario is the extension of Scenario 2 in the large environment with five teams of agents, totaling ten agents, and two distinct goals. The objective is to validate whether A5 outperforms A1 in the large environment.
\end{enumerate}

\subsection{Results and Discussion} \label{sec:r_n_d}

\begin{figure}[!t]
    \centering
    \includegraphics[width=0.85\linewidth]{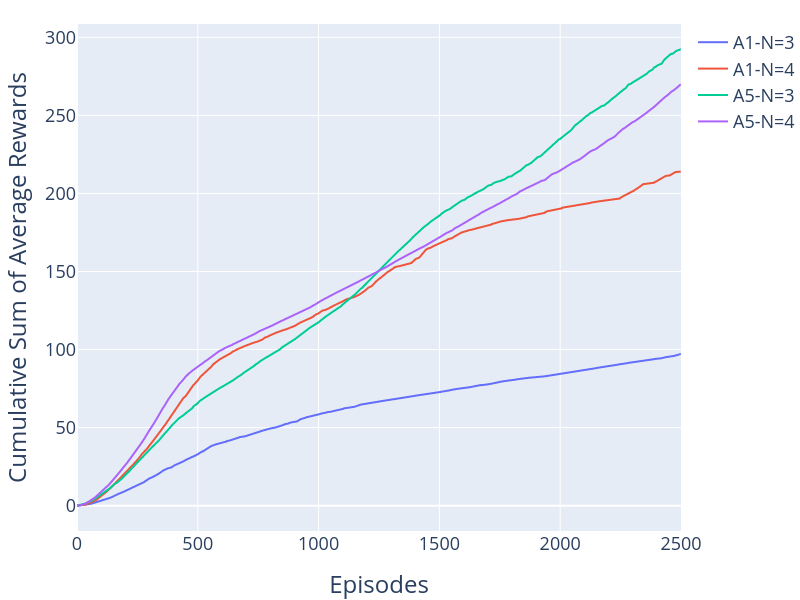}
    \caption{Comparison between A1 and A5 in Scenario 2.}
    \label{fig:s2}
    \vspace{-6pt}
\end{figure}

\begin{table}[!t]
    \centering
    \begingroup
    \scriptsize
    \renewcommand{\arraystretch}{1.5}
    \begin{tabular}{|c|c|c|c|c|c|}
        \hline
        \textbf{Types of Agent} & $\boldsymbol{N}$ & \textbf{Agent 1} & \textbf{Agent 2} & \textbf{Agent 3} & \textbf{Agent 4} \\
        \hline
        A1 & 3 & 243 $\pm$ 93 & 178 $\pm$ 103 & 166 $\pm$ 92 & N/A \\
        A5 & 3 & 171 $\pm$ 93 & 136 $\pm$ 93 & 166 $\pm$ 92 & N/A \\
        A1 & 4 & 243 $\pm$ 93 & 178 $\pm$ 103 & 166 $\pm$ 92 & 87 $\pm$ 91 \\
        A5 & 4 & 171 $\pm$ 93 & 136 $\pm$ 93 & 176 $\pm$ 95 & 101 $\pm$ 92 \\
        \hline
    \end{tabular}
    \endgroup
    \caption{The average number of steps taken by each agent during successful episodes in Scenario 2.}
    \label{tab:s2}
\end{table}

\subsubsection{Scenario 1}

\blue{The overall performance of the system with agents restricted in both collaboration and observation ranges (A5) outperforms all other agent types (see also Figure \ref{fig:s1}). The results also show that independent agents (A1) outperform collaborative agents without any restrictions (A2). Without restrictions on individual goal awareness, agents can learn irrelevant information shared by agents with different goals at each time step. This can lead to sub-optimal action selection and requiring more steps for task completion (see Table \ref{tab:s1}).}

\blue{To address this issue, we designed A3 and A4. Introducing observation ranges for agents (A3) improves performance, and agents tend to take fewer steps to complete tasks compared to A1 (see Table \ref{tab:s1}). This improvement likely results from observation ranges reducing the time steps where agents learn irrelevant information from others with different goals. Without observation ranges, it becomes essential to filter irrelevant information by restricting collaboration to agents with the same individual goal (A4). Agents are grouped into teams if they share the same individual goal. Our experiments revealed three key insights: (i) A4 outperforms A1, A2, and A3; (ii) agents took the fewest steps to complete tasks compared to other agent types; and (iii) the overall system performance converged the fastest. However, agent performance declined after convergence. This suggests that while a team of agents can learn quickly, it can overfit without observation ranges.}

\blue{To address this issue, we designed A5. Although A5 underperforms A4 during the first 1300 episodes, it helps avoid the overfitting problem in the long run. The results show that the overall system performance continues to improve over the course of 2500 episodes (see Figure \ref{fig:s1}).}

\subsubsection{Scenario 2}

\blue{When introducing an additional agent to the environment to establish two teams, the overall performance of the system with A5 still outperforms that of A1 (refer to Figure \ref{fig:s2}). Since the performance of Agents 1 and 2 remains unchanged, we focus on analyzing the performance of Agents 3 and 4 in this scenario. When operating as independent agents (A1), Agent 4 surpasses Agent 3, completing tasks with fewer steps (see also Table \ref{tab:s2}). This may be attributed to Agent 4's closer position to the goal compared to Agent 3. 
However, when Agents 3 and 4 engage in communication and coordination during task execution (A5), the performance of Agent 4 gradually declines. This can be because the low performance of Agent 3 negatively affects Agent 4 during coordination. Furthermore, Figure \ref{fig:s2} illustrates that system performance with four agents grows faster than with three agents during the first 1300 episodes. This highlights the importance of mitigating the impact of poorly performing agents when scaling our approach to include more agents.}

\subsubsection{Scenario 3}

\begin{figure}[!t]
    \centering
    \includegraphics[width=0.85\linewidth]{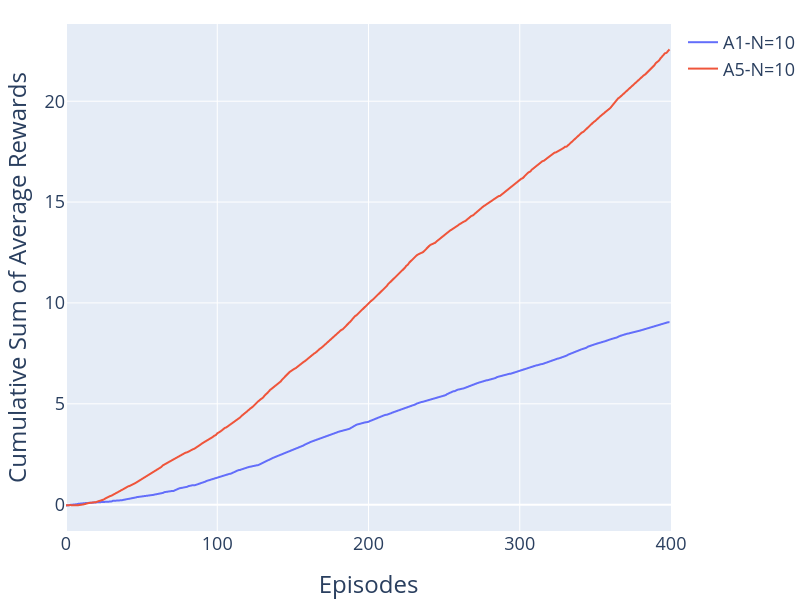}
    \caption{Comparison between A1 and A5 in Scenario 3.}
    \label{fig:s3}
    \vspace{-16pt}
\end{figure}

\blue{Figure \ref{fig:s3} illustrates that A5 outperforms A1 even in the larger environment with more agents. During our experiments, we observed that a batch size of 64 was insufficient for agents to effectively learn from their experiences in such large environment. Therefore, we increased the batch size to 256. In addition, a time limit of $\mathcal{T} = 400$ was inadequate for some agents to reach their goals, often resulting in negative episodic rewards even for successful episodes. To address this, we increased $\mathcal{T}$ to 2500 and reduced the number of episodes $\mathcal{E}$ to 400, ensuring that $\mathcal{E} \times \mathcal{T} = 10^6$ 
The results also shows that the success rate of agents with A5 improves by 20\% compared to those with A1. Furthermore, agents with A5 tend to take fewer steps to complete tasks than those with A1.}

\section{Conclusion and Future Work} \label{sec:concl}
We proposed a novel fully decentralized Multi-Agent Reinforcement Learning (MARL) approach that enables goal-aware coordination agents. Applied to a multi-agent navigation task in complex grid environments with obstacles, our method outperformed non-collaborative agents by achieving faster task completion. Notably, it maintained strong performance even as the number of agents increased, demonstrating scalability in decentralized settings. For future work, we aim to address the negative impact of poorly performing agents on overall system performance during scaling. Additionally, we plan to evaluate the robustness of our approach in real-world scenarios, such as multi-drone search and rescue missions. Given its applicability across domains, future research will also explore domain-specific reward shaping strategies.

\bibliographystyle{ieeetr}
\bibliography{references.bib}

\appendix

\section*{Appendix 1: Multi-Agent Reinforcement Learning Algorithm with Decentralized Coordination} \label{sec:app_1}

\begin{algorithm}[!ht]
    \caption{Multi-Agent Reinforcement Learning Algorithm with Decentralized Coordination} \label{algo:dfm}
    \begin{algorithmic}[1]
        \State Randomly initialize critic network per agent  $Q_{i}$ with $\theta_{i}^Q$ and its target network $\theta_{i}^{Q'} \leftarrow \theta_{i}^Q$
        \State Randomly initialize actor network per agent $\mu_{i}$ with $\theta_{i}^\mu$ and its target network $\theta_{i}^{\mu'} \leftarrow \theta_{i}^{\mu}$
        \State Initialize relay buffer per agent $\{ \mathcal{B}_{i} \}_{i=1}^{n}$
        \For{episode = 1, $M$}
            \State Initialize observation state per agent $s^{1}_{i}$
            \For{t = 1, T}
                \State Get observations of each agent $\{ \mathcal{O}^{t}_{i} \}_{i=1}^{n}$
                
                \For{each agent $i$}
                    \If{$i$ is not terminated}
                        \State Identify other agents $\{ j \}^{n}_{j \neq i}$ where $(x^{t}_{j}, y^{t}_{j}) \in \mathcal{O}^{t}_i \land g_{i} = g_{j}$
                        \State Update $\theta_{i}^Q$ and $\theta_{i}^{\mu}$ according to the coordination strategy
                    \EndIf
                \EndFor
                \State Identify agents $\{ i \}$ that have not been terminated
                \For{each agent $i$}
                    \State Select action $a^{t}_{i}$ according to the current policys
                    \State Execute action $a^{t}_{i}$ and observe reward $r^{t}_{i}$ and the new state $s^{t+1}_{i}$
                    \State Store the transition $\left( s^{t}_{i}, a^{t}_{i}, r^{t}_{i}, s^{t+1}_{i}  \right)$ in $\mathcal{B}$
                    \State Sample a random minibatch of $N$ transitions $\left( s^{h}_{i}, a^{h}_{i}, r^{h}_{i}, s^{h+1}_{i} \right)$ from  $\mathcal{B}$
                    \State Set $y^{h}_{i} = r^{h}_{i} + \gamma Q'\left( s^{h+1}_{i}, \mu'\left( s^{h+1}_{i} | \theta^{\mu'}_{i} \right) | \theta^{Q'}_{i} \right)$
                    \State Update critic by minimizing the loss (using $y^{h}_{i}$ and $\theta_{i}^{Q}$ for Equation 3)
                    \State Update the actor policy using the sampled policy gradient and Equation 6 
                    \State Update target networks:
                    \begin{align*}
                        & \theta^{Q'}_{i} \leftarrow \tau\theta^{Q}_{i} + (1 - \tau)\theta^{Q'}_{i} \\
                        & \theta^{\mu'}_{i} \leftarrow \tau\theta^{\mu}_{i} + (1 - \tau)\theta^{\mu'}_{i}
                    \end{align*}
                    \If{agent $i$ reaches its goal}
                        \State Terminate $i$
                    \EndIf
                \EndFor
            \EndFor
        \EndFor
    \end{algorithmic}
\end{algorithm}

\blue{Algorithm \ref{algo:dfm} outlines our approach for training agents in fully decentralized settings. For experimental purposes, we follow steps similar to the Deep Deterministic Policy Gradient (DDPG) algorithm \cite{lillicrap2015continuous}. In action selection (Line 16), we replace the Ornstein-Uhlenbeck process \cite{uhlenbeck1930theory} used in DDPG with a multinomial sampling process. Consequently, entropy regularization terms are incorporated into the actor loss estimation (Line 22). Based on our empirical experiments, this combination enhances the agents' exploration process. Importantly, the novelty of our approach lies in the communication and coordination strategy that incorporates individual goal awareness (described in Lines 7-13). It is important to note that if an agent cannot identify its peers, its learning weights will not be updated (Line 11) during the collaboration session.}

\begin{figure}[!t]
    \centering
    \includegraphics[width=\linewidth]{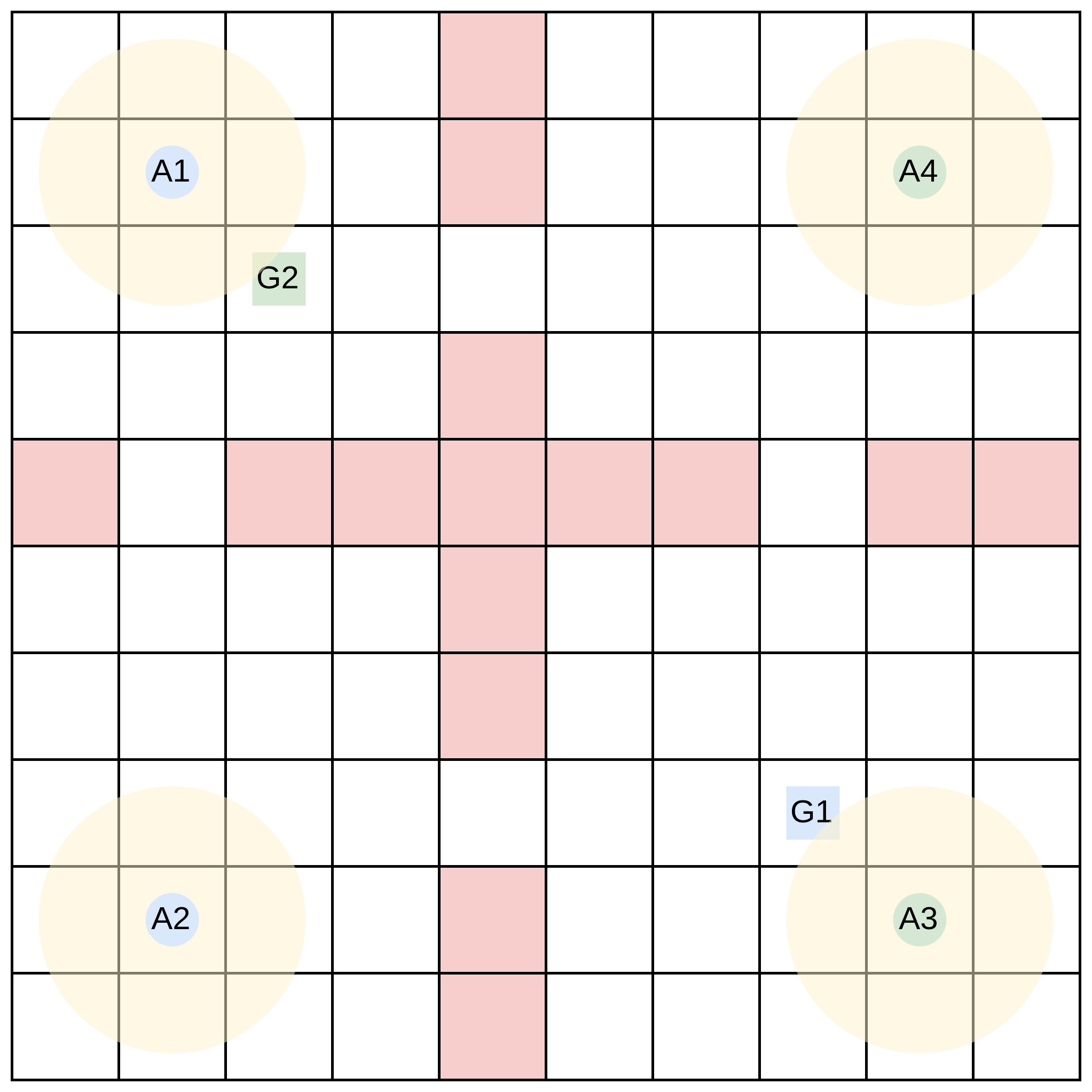}
    \caption{An overview of the small environment for our experiments. In this scenario, A1 and A2 pursue G1, while A3 and A4 pursue G2.}
    \label{fig:env1}
\end{figure}

\begin{figure}[!ht]
    \centering
    \includegraphics[width=\linewidth]{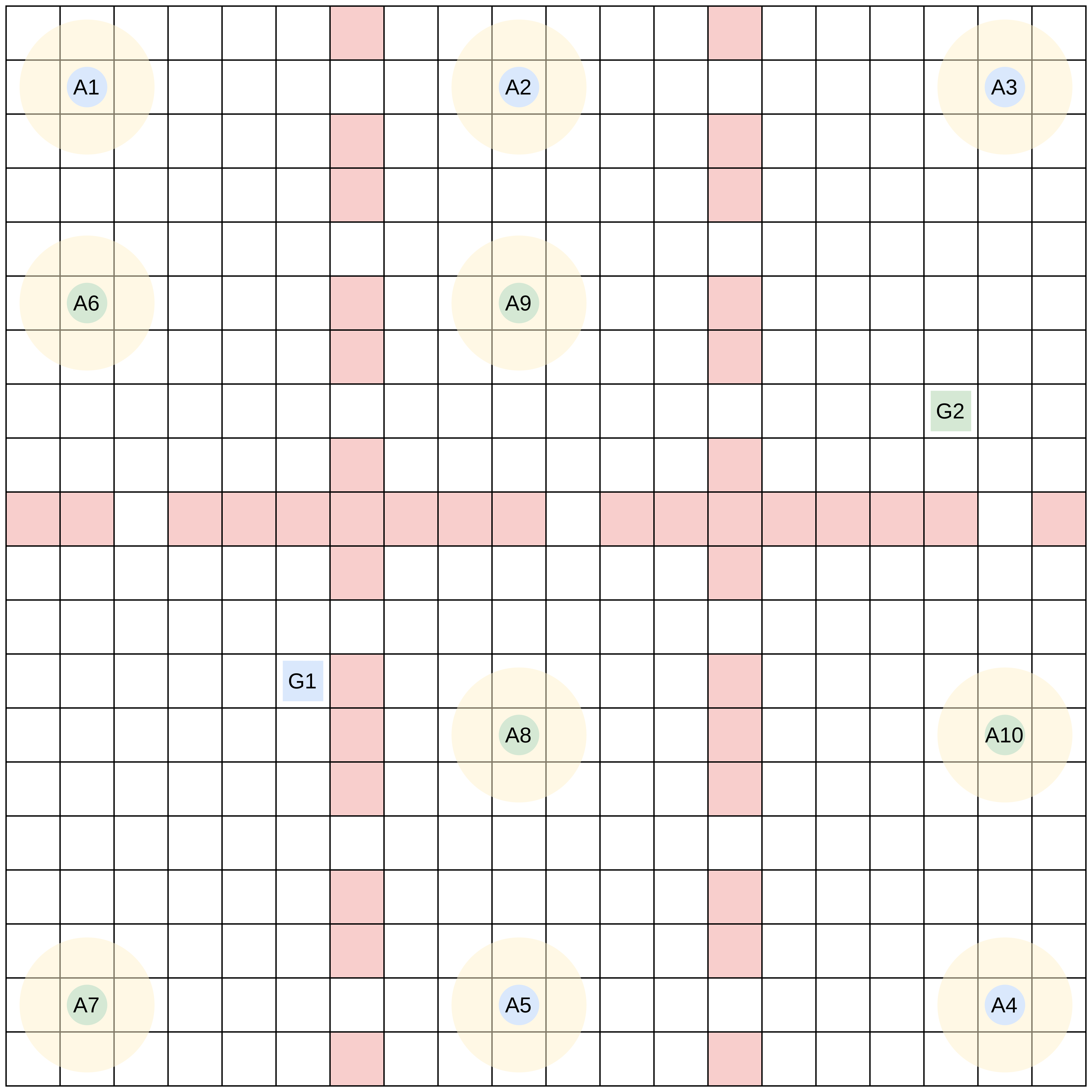}
    \caption{An overview of the large environment for our experiments. In this scenario, A1 to A5 pursue G1, whereas A6 through A10 pursue G2.}
    \label{fig:env2}
\end{figure}

\section*{Appendix 2: Environments} \label{sec:app_2}

\blue{We designed two challenging 2D grid environments with dimensions of $10 \times 10$ (small) and $20 \times 20$ (large) (see Figures \ref{fig:env1} and \ref{fig:env2}, respectively). Each environment includes three main entities: agents (represented as circles), agent goals (depicted as squares), and obstacles (shown as filled red cells). In addition, a yellow circle indicates the observation range around each agent.}

\blue{The primary objective of our design is to evaluate the impact of agent interaction within a team on improving task completion. To this end, the environment is structured as multiple interconnected rooms. Each room contains several doors that allow agents to move to adjacent rooms and includes at least one agent. In addition, a room may contain a goal specific to a team of agents. If an agent starts in this type of room, its goal is always different from the goals in the room. This aims to increase the environment's complexity. In the small environment, each room has a single door, enabling agents to transition between adjacent rooms on either side (see Figure \ref{fig:env1}). In contrast, the large environment introduces multiple doors for horizontal navigation between rooms to reduce the environment's complexity (see Figure \ref{fig:env2}).}

\end{document}